\documentclass[%
 reprint,
 amsmath,amssymb,
 aps,
]{revtex4-2}

\usepackage{graphicx}% Include figure files
\usepackage{dcolumn}% Align table columns on decimal point
\usepackage{bm}% bold math
\usepackage{siunitx}
\usepackage{hyperref}% add hypertext capabilities
\usepackage{color}

\begin{document}
\preprint{APS/123-QED}

\title{Laser frequency offset locking at 10-Hz-level instability using hybrid electronic filters}

\author{Vyacheslav Li}
%\email{vyacheslav.li@ist.ac.at}
%\affiliation{Institute of Science and Technology Austria, Am Campus 1, 3400 Klosterneuburg, Austria}

\author{Fritz Diorico}
%\email{fritz.diorico@ist.ac.at}
%\affiliation{Institute of Science and Technology Austria, Am Campus 1, 3400 Klosterneuburg, Austria}

\author{Onur Hosten}
\email{onur.hosten@ist.ac.at}
\affiliation{Institute of Science and Technology Austria, Am Campus 1, 3400 Klosterneuburg, Austria}
\date{\today}

\begin{abstract}    
    Lasers with well controlled relative frequencies are indispensable for many applications in science and technology. We present a frequency offset locking method for lasers based on beat frequency discrimination utilizing hybrid electronic LC filters. The method is specifically designed for decoupling the tightness of the lock from the broadness of its capture range. The presented demonstration locks  two free running diode lasers at 780 nm with a \SI{5.5}{\giga\hertz} offset. It displays an offset frequency instability below \SI{55}{\hertz} for timescales in excess of \SI{1000}{\second} and a minimum of \SI{12}{\hertz} at \SI{10}{\second} averaging, outperforming the best reported instabilities of methods based on beat frequency discrimination. The performance is complemented with a \SI{190}{\mega\hertz} lock capture range, a tuning range of up to \SI{1}{\giga\hertz}, and a frequency ramp agility of \SI{200}{\kilo\hertz/\micro\second}.
\end{abstract}
\maketitle
\section{Introduction}

The ability to generate laser beams with well controlled and stable relative frequencies enables applications in many scientific and technological areas. Precision optical metrology using optical frequency combs~\cite{Udem2002}, laser cooling and trapping of atoms~\cite{Metcalf1999}, coherent manipulation of their internal~\cite{Fleischhauer2005} as well as their motional degrees of freedom~\cite{Kasevich1991,Hu2017}, and coherent manipulation of solid-state quantum systems~\cite{Santori2006,Bodey2019} are examples relying on good relative frequency control. The capability of controlling such systems makes possible precision applications like atomic clocks~\cite{Ludlow2015} and atom interferometers~\cite{Tino2014} for geodesy, inertial navigation and tests of fundamental physics~\cite{Szigeti2021}, and makes possible many quantum information processing applications.

%and coherent optical communications~\cite{Harrison1989,Al-Taiy2014}

A number of general techniques exist for generating laser tones with fixed relative frequencies. These include splitting a single laser beam  and shifting its frequency with acousto-optic~\cite{Buchkremer2000} or electro-optic modulators (EOM)~\cite{Johnson2010} -- reaching offsets from tens of MHz to tens of GHz; locking separate lasers to different resonances of optical cavities~\cite{Day1992} or atomic systems~\cite{Nakayama2006,Bell2007} -- reaching from GHz to hundreds of THz; and directly stabilizing the frequency offset between lasers by utilizing the interference beat note between the optical tones on a photodetector -- offering the largest range of applicability while keeping lasers spectrally pure. This last technique will form the basis of the method developed in the present work. 

A direct beat note between two lasers can be detected up to tens of GHz on fast photodiodes, and offsets up to a couple of hundred GHz is achievable by beating one of the lasers with a high-order sideband of another laser resulting from its phase modulation with an EOM~\cite{Greve2021}. All the way up to octave-spanning THz offset frequencies can be achieved by utilizing the beat notes of two lasers with different spectral lines of an optical frequency comb~\cite{Nicolodi2014}, enabling optical clock frequency comparison.

Phase locking of two lasers with the aid of an optical beat note~\cite{Santarelli1994,Cacciapuoti2005} in principle produces the tightest frequency lock, reaching below sub-mHz instability levels. This optical phase locked loop (OPLL) method however places the most stringent demands on feedback stabilization bandwidths, requiring it to be appreciably larger than the linewidth of the lasers. This makes OPLLs especially challenging with semiconductor diode laser systems which possess linewidths from several hundred kHz to several tens of MHz. Nevertheless, many applications do not require such stringent locking stabilities, in which case laser frequencies can more easily be steered and locked using purely beat frequency discrimination. 

Among typically utilized discriminators, simple delay-line based architectures~\cite{Schunemann1999,Hisai2018} have been demonstrated to reach 1 kHz frequency difference instability, but they suffer from the technical complications brought by periodically repeating locking points. Digital counting architectures~\cite{Hughes2008} have been demonstrated to reach 300 Hz instability, but suffer from spurious digital noise and implementation complexity. Architectures relying on integrated frequency-to-voltage conversion circuits have reached instabilities in the 30-100 Hz range, but they also suffer from implementation complexities that come about overcoming the limited frequency range ($\sim$1 MHz) of such chips. Similarly, side-of-filter type architectures~\cite{Ritt2004,Puentes2012,Strauss2007} have also been demonstrated to reach down to the 100 Hz instability regime. Basic implementations of these architectures typically trade-off locking stability with the initial capture range of the lock, posing a general additional limitation.

Here we develop a new frequency discrimination method based on hybrid LC filters for laser offset locking. The method is developed for the goals of obtaining a single lockable point with simultaneous wide-tunability, high locking stability, and a broad capture range. The implemented frequency discriminator (FD) generates an error signal with a 1.8-MHz-wide steep region and broad tails that extend out to hundreds of MHz for ensuring a large capture range. The all-analog design of the offset locking circuitry ensures low-noise operation. An offset frequency tunability of \SI{1}{\giga\hertz} is incorporated by heterodyning the beat signal with a tunable microwave local oscillator (LO) generated from a direct digital synthesizer (DDS).

We demonstrate the developed method on two 780-nm miniature external-cavity diode lasers (ECDL) with $\sim$500~kHz observed free-running linewidts. We stabilize the frequency offset between the two lasers at around \SI{5.5}{\giga\hertz}, and reach a frequency locking instability of \SI{12}{\hertz} at \SI{10}{\second} averaging time. This value corresponds to a fractional instability of $3\times10^{-14}$ when scaled to the laser frequency. The long term instability remains below \SI{55}{\hertz} for more than \SI{1000}{\second}. To our knowledge, these results constitute the lowest beat-note-based offset locking instability published in absence of optical phase locking. Beyond its exceptional stability, the laser frequency offset is tunable at rates exceeding \SI{200}{\kilo\hertz/\micro\second} by frequency ramping the DDS generating the LO. We note that the use of a simple LC circuit for laser offset locking was reported recently also in Ref.~\cite{Cheng2017}, but the operation was limited to $\sim$2~kHz instability levels. Below, we first detail our scheme and then describe the experimental characterization of its performance.

\section{The scheme}
\label{method}

\begin{figure}[t]
\centering
	\includegraphics[width=1\columnwidth]{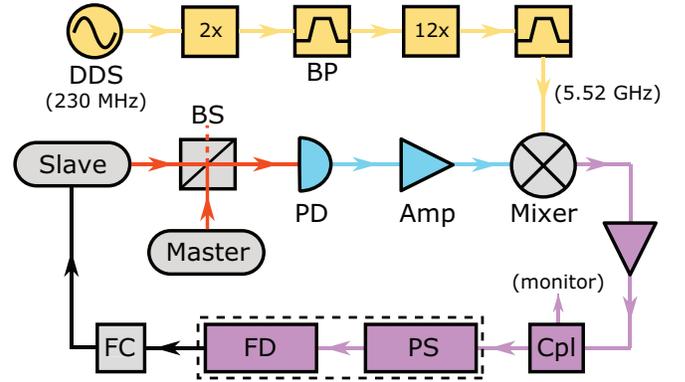}
	\caption{Block diagram of the frequency offset locking setup. The master and slave lasers form a beat note on a fast photodetector (PD). The signal at $\thicksim$\SI{5.5}{\giga\hertz} frequency difference (blue) is amplified and mixed down with a local oscillator (LO) chain (yellow) that starts with a direct digital synthesizer (DDS). The down-converted output (purple) of the mixer is in the \SI{}{\mega\hertz} range, and is amplified to the saturation point. A directional coupler (Cpl) is used to monitor the beat note. The signal goes through an active power stabilization (PS) unit to stabilize the power at the input of the frequency discriminator (FD). The feedback controller (FC) then manipulates the slave laser current to stabilize its frequency at a well defined offset from the master. BP: bandpass filter, BS: beam splitter. Additional amplifiers in the LO chain are omitted from the diagram. Part list; PD: Thorlabs DX12CF, Amp: Mini-circuits ZX60-83LN-s+ and ZX60-43-s+, Mixer: ZX05-83-s+, Cpl: ZFDC-10-1-s+, 2x: ZX90-2-13-s+, 12x: ZX90-12-63-s+, DDS: Artiq Urukul, Master$\&$Slave: Sacher Micron. }
	\label{fig:block_diagram}
\end{figure}

Figure~\ref{fig:block_diagram} shows a block diagram of the laser offset frequency locking scheme. Small portions of a master laser and a slave laser are superimposed on a 12-GHz-bandwidth photodetector, and the resulting beat note is amplified and heterodyned with a microwave LO. The frequency down-converted signal is further amplified, and its power is actively stabilized. This signal is then fed into the hybrid electronic FD to generate a baseband error signal that has opposite polarities on opposite sides of the desired frequency offset point. A home-built feedback controller is then used to stabilize the slave laser frequency at this controllable offset point with respect to the master laser. The down-conversion process that brings the beat signal from the GHz regime to the MHz regime eases further electronic processing of the signal. At the same time, it enables the wide-range tunability of the frequency offset through induced changes in the LO frequency.

\begin{figure}
\centering
	\includegraphics[width=1\columnwidth]{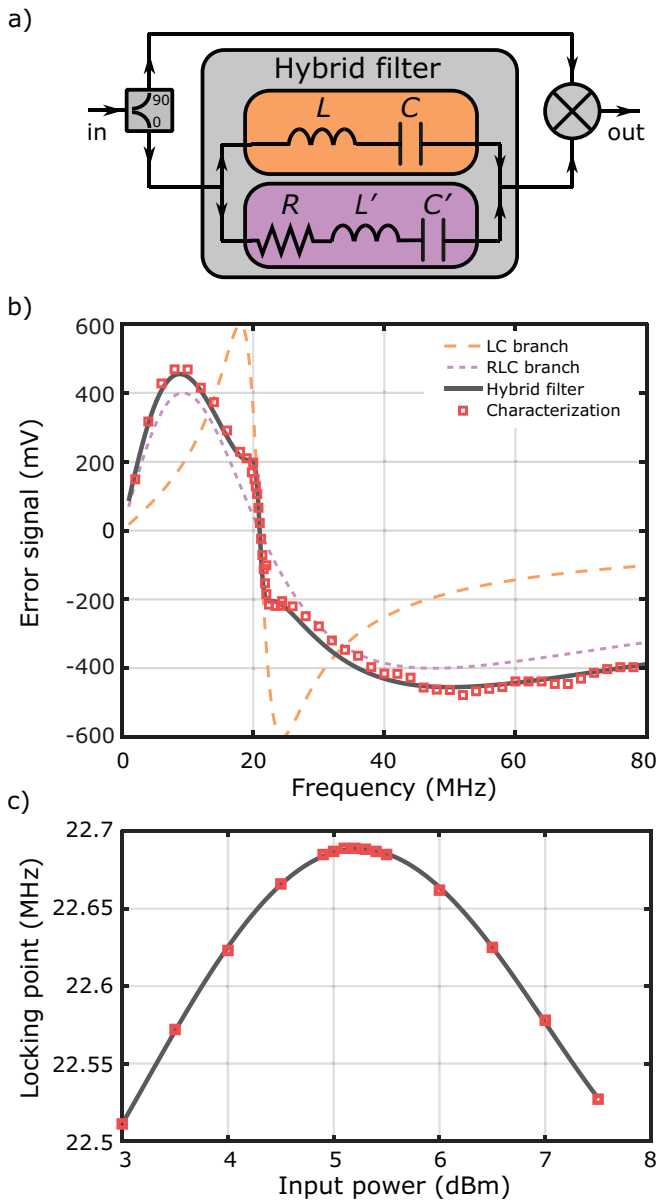}
	\caption{(a) Schematic drawing of the FD. The hybrid filter (gray box) is composed of two branches: one LC filter (orange box) and one RLC filter (purple box). It is part of a Mach-Zehnder-like interferometer composed of a splitter (Mini-circuits PSCQ-2-51W+1) and a mixer (RPD-1+1 phase detector). The surface-mount component values are $L=2.2~\mathrm{\mu H}$, $C= 20~\mathrm{pF}$ and $R=50~\Omega$, $L'=470~\mathrm{nH}$, $C'= 91~\mathrm{pF}$. (b) Expected and observed frequency ($\omega/2\pi$) responses of the FD given by the imaginary part of the transfer function $H_{im}(\omega)$, corresponding to the error signal. The theoretical response curves are scaled in magnitude to fit the experimentally characterized response (red squares). Contributions of individual branches are also explicitly indicated. (c) Zero-crossing frequency of the generated error signal as a function of input signal power at the FD (red squares), and a polynomial fit (solid line). All characterizations were carried out with a frequency stepped tone from a function generator.}
	\label{fig:exp_char}
\end{figure}

The FD (Fig.~\ref{fig:exp_char}a) is the key component that generates the error signal with a tight resonance and a wide capture range. It is a Mach-Zehnder-like interferometer with a specially designed hybrid filter in one of the two arms. The input signal $\sim sin(\omega t+\phi_0)$ is distributed into the arms with a 2-way \SI{90}{\degree}-power-splitter. Here $\omega$ is the angular frequency and $\phi_0$ is an arbitrary phase. The arm with the hybrid filter contains the waveform $sin(\omega t + \phi_0)\equiv Re[i\,e^{-i(\omega t + \phi_0)}]$ while the other arm contains the $cos(\omega t + \phi_0)$ waveform. The filter transforms the incoming waveform with its transfer function $H(\omega)=H_{re}(\omega) +i H_{im}(\omega)$, resulting into the transmitted signal $Re[H(\omega)\,i\, e^{-i(\omega t + \phi_0)}]= H_{re}(\omega)\,sin(\omega t + \phi_0) - H_{im}(\omega)\,cos(\omega t + \phi_0)$. This is then combined on a frequency mixer with the $cos(\omega t + \phi_0)$ waveform coming from the other arm to result into the baseband signal proportional to $H_{im}(\omega)$ at the mixer output, which is used as the error signal (Fig. \ref{fig:exp_char}b). 

The hybrid LC filter design eliminates the usual trade-off between lock tightness and capture range. The LC circuit contained in the upper branch of the hybrid filter (Fig.~\ref{fig:exp_char}a) possesses a narrow resonance for tighter locking, and the one in the lower branch possesses a broad resonance for obtaining a wide capture range. The resulting hybrid filter thus inherits both properties. The resonance frequencies of the two filter branches must match to ensure the symmetry of the error signal in the vicinity of its zero-crossing. The location of the zero-crossing determines the exact frequency offset between the lasers, since the feedback loop stabilizes the error signal at the zero value. In our design, a resonance frequency of \SI{22.6}{\mega\hertz} was chosen such that it is high enough to obtain a sufficiently large capture range, but low enough to facilitate an easy circuit implementation. Note that although the feedback loop always locks the signal seen by FD at 22.6 MHz, the actual laser offset frequency is given by the sum of this value and the frequency of the tunable LO.  

The frequency of the slave laser is controlled through its injection current. A home-built analog feedback loop utilizing the error signal generated by the FD controls this current with a unity-gain bandwidth of $\sim$\SI{100}{\kilo\hertz}. The loop contains two intergrators with the second one kicking in for frequencies below 20kHz, giving rise to a $1/\omega^2$ feedback strength down to DC. The implementation of the feedback controller was carried out with particular attention to ground loops, which is typically a major source of stability degradation in many applications. The circuit board was designed such that all inputs and outputs are carried out differentially (interchangeably using AMP03 and LT1167 ICs), cutting ground connections between various devices, hence eliminating ground voltage shifts across the circuit board. In the implemented frequency feedback, the FD output was directly connected to the feedback controller after 50~$\Omega$ termination and some attenuation for loop gain tuning. No lowpass filtering was utilized to eliminate the harmonics of the beat note that the FD additionally generated. Such filtering was observed to shift the locking point -- presumably due to back-reflections into the FD -- but did not seem to noticeably affect the locking stability.

Real-world analog mixers can develop an input-power dependent residual DC-offset in their outputs, even when no mixing to DC is expected in the ideal case~\cite{Rubiola2006}. Thus, power fluctuations at the FD -- containing a mixer -- have the potential to turn into small additive fluctuations in the generated error signal. Since the feedback controller would mistake these for frequency offset fluctuations, power drifts and fluctuations could lead to degradation of the actual frequency offset stability. In order to suppress such effects, we first reduce power fluctuations by saturating the amplifiers leading to the FD (Fig.~\ref{fig:block_diagram}), and further utilize additional active feedback stabilization to fix the beat note power seen by the FD. This feedback circuitry (not shown) splits (Mini-circuits PSC-2-1+) the signal into two paths, and self-mixes (SRA-1-1+) the sample branch to effectively measure the power, and sends the other branch to the FD. A voltage variable attenuator (PAS-3+1) situated before the splitter then stabilizes the power by means of a low-bandwidth feedback loop utilizing the power measurement. Note, that the self mixing is carried out with power levels at least 10 dB below the mixer input-saturation point. The PCB circuit housing the FD and the power control circuitry (dashed box in Fig.~\ref{fig:block_diagram}) is operated with a nominal input power of 17 dBm, and remains identically functional down to 11 dBm. The amplifier saturation and the active RF power control provides a buffer for potential beat note power degradations over time (up to 10 dB) without compromising operational performance.

\section{Experimental characterization}
\label{section:exp_char}

The error signal generated by the FD was characterized by applying a single tone from a function generator while stepping its frequency to map the response (Fig.~\ref{fig:exp_char}b). The sharp region of the error signal was measured to be \SI{1.8}{\mega\hertz} in agreement with the design parameters. The lock capture range was empirically found to be around \SI{190}{\mega\hertz}, ranging from 45 MHz below the unique locking point to 145 MHz above. The error signal tails extend further than this (Fig.~\ref{fig:exp_char}b), but the noise on the error signal -- originating from finite laser linewidths -- hinders lock acquisition for initial conditions too far from the locking point. In free-running operation, the employed laser systems do not drift more than $\sim$50 MHz, justifying the choice of the design parameters leading to the observed capture range.

The effect of the non-ideal behavior of the FD mixer was characterized by measuring the error signal zero-crossing frequency as a function of the RF input power to the FD unit. Empirically, a `turning point' was identified around \SI{5.2}dBm  (Fig.~\ref{fig:exp_char}c) where the zero-crossing frequency was insensitive to the input power. These measurements were carried out with a single tone from a function generator. The exact power corresponding to the `turning point' however appears to be linewidth dependent, and needs to be fine tuned in locked operation by observing the resulting mean frequency offset between the lasers. With saturated amplifiers and additional active power stabilization at the `turning point', the stability was no longer limited by power fluctuations. 

%The location of the zero-crossing determines the exact frequency offset between the lasers, since the feedback loop stabilizes the error signal at the zero value.

\begin{figure}[t]
\centering
	\includegraphics[width=1\columnwidth]{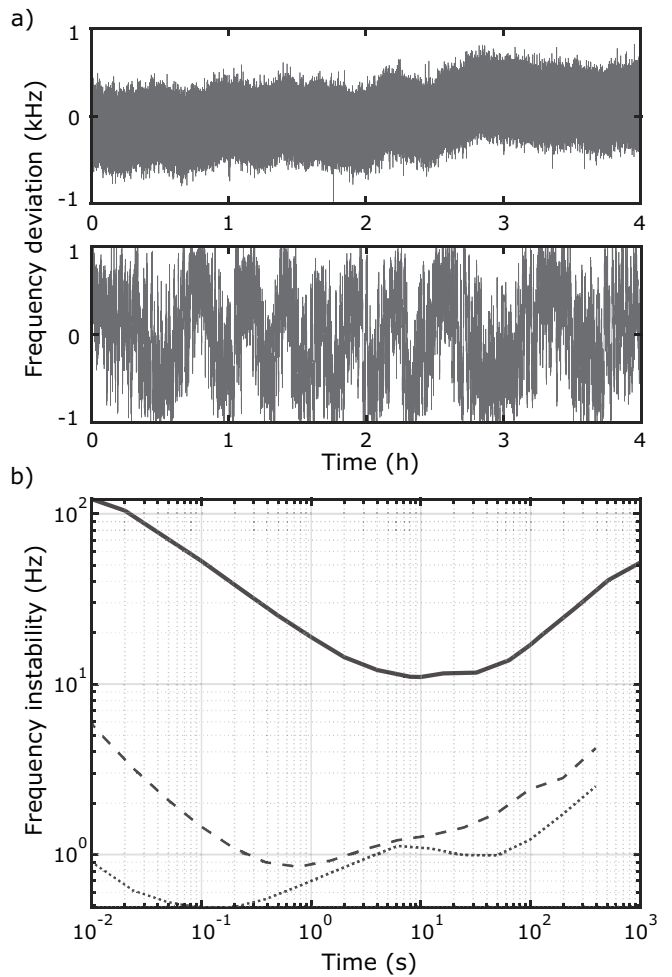}
	\caption{(a) Frequency deviation of the monitored beat frequency with (upper) and without (lower) power stabilization. Counter gate time: \SI{0.01}{\second}. (b) Allan deviation of the observed beat frequency (solid line) for the power stabilized case. A minimum of \SI{12}{\hertz} at \SI{10}{\second} averaging is reached, and the instability remained below \SI{55}{\hertz} for timescales exceeding \SI{1000}{\second}. The dashed line is the instability that would be naively inferred from the recorded residual error signal while the lock is engaged. The dotted line indicates the electronic noise floor inferred by blocking the laser light.}
	\label{fig:data}
\end{figure}

The evaluation of the achieved offset locking stability was carried out by recording the beat note frequency through the 22.6 MHz monitor output (Fig.~\ref{fig:block_diagram}) using a frequency counter (SRS FS470). Sample time traces with and without power stabilization are shown in Figure~\ref{fig:data}a, displaying the utility of power stabilization. For the power stabilized case, the mean frequency remains within a \SI{200}{\hertz} window over the course of a day. The characterized frequency offset instabilities at different time scales for this case are obtained by calculating the Allan deviation (Fig.~\ref{fig:data}b) of the recorded beat note data. The instability remains below \SI{55}{\hertz} for more than \SI{1000}{\second} and has a minimum value of \SI{12}{\hertz} at \SI{10}{\second} timescale. The measured lock tightness inferred from the closed loop error signal, and the limitations induced by detection noise measured in open loop with the lasers blocked (Fig.~\ref{fig:data}b), both contribute at least an order of magnitude lower than the observed frequency offset instability. The origin of the dominant contribution to the instability is currently unclear.

The agility of user inducible laser frequency offset changes is related to the bandwidth of the feedback loop controlling the slave laser frequency. In the current setup, continuous ramp rates of \SI{200}{\kilo\hertz/\micro\second} were achievable robustly. The range of instantaneous frequency jumps on the other hand were limited by the capture range of 190 MHz. Such step jumps settle towards the setpoint with a time constant of order 10 $\mu s$, determined by the inverse of the lock bandwidth. The frequency offset can be tuned within a range of 1 GHz, limited by the bandpass filters utilized in the microwave LO generation chain. In principle, given a widely tunable LO source, the only limitation to the tuning range is brought by the 12-GHz bandwidth of the photodetector (Fig.~\ref{fig:block_diagram}).

\section{Discussion}
\label{conclusion}

We designed and implemented a frequency offset locking method based on hybrid electronic LC filters. We stabilized two 780-nm miniature ECDL lasers at a \SI{5.5}{\giga\hertz} offset from each other. With attention to details like input power stabilization at the FD and utilization of ground loop elimination techniques for the feedback controller, we reached an unprecedented difference frequency locking instability of 12 Hz at 10 s averaging. Some inherent problems of earlier filter-type implementations were thus circumvented with these additional techniques. The hybrid nature of the FD decoupled the problem of the capture range from that of lock tightness. The attained stability, the capture range and the wide tunability make this scheme very appealing for applications where the relative frequencies need to be stabilized and tuned on demand. In this work, the sharp filter region extending 1.8 MHz was chosen to make the linear range of the error signal larger than the laser linewidths, but in principle it can be made substantially narrower. We observed that both the sharpness of the resonance (by trying a different filter) and the reduction of power fluctuations increase the stability.

The specific system in this work was developed to frequency stabilize the cooling and re-pumper beams for laser cooling and trapping of $^{87}$Rb atoms. In the final configuration, the master laser would be locked to the $3 \leftrightarrow 4'$ transition of the $^{85}$Rb D2 line via modulation transfer spectroscopy (not discussed). Two slave lasers (only one discussed) would then be locked at 5.5 GHz and 1.2 GHz offsets from the master laser using the described methods. The slave lasers will respectively drive the $1 \leftrightarrow 2'$ re-pumping and the $2 \leftrightarrow 3'$ cooling transitions of $^{87}$Rb. The agile tunability is intended for the required ramping of the cooling beam detuning. The achieved stability levels are excessive for the current application, nevertheless, the methods remain general, and can be used for other demanding applications.

In this work, there was no differential linewidth narrowing for the lasers since the lock bandwidth is appreciably smaller than the laser linewidths. In a system that employs either smaller linewidth lasers or a larger bandwidth feedback, the relative linewidth can be collapsed to very small values. This will have an immediate impact on the short term stability, pulling the minimum of the Allan deviation to lower values. Additionally, this could also improve the long term stability: notice that the line center is locked with a 12-Hz instability, but the linewidth is several hundred kHz. Any changes in the beat line shape could alter the effective line center seen by the FD. Such line shape changes could originate from the lasers themselves, or the time variation of residual back reflections in the FD that are known to be frequency dependent. This could cause discrepancies between what is measured by the FD and what is measured by the counter used to verify the stability. A more systematic analysis of the stability limitations can be carried out in the future with a narrow line beat note. Lastly, although saturating the amplifiers improves robustness and helps increase long term stability, this process inevitably generates higher harmonics. These additional RF tones could potentially result in residual time-varying DC signals at the output of the FD. We have not investigated such mechanisms that might contribute to the observed instabilities.

As a future application, in addition to its use for purely frequency locking, given its large capture range and stability, the demonstrated offset locking method can be combined with OPLLs to ease acquisition of lock and improve the robustness against transients causing permanent loss of lock~\cite{Cheng2017,Seishu2019}. This could replace the more complicated and noisy digital phase-frequency detectors utilized in analog+digital low-noise OPLL systems~\cite{Cacciapuoti2005}.

\begin{acknowledgments}
This work was supported by IST Austria. The authors thank Yueheng Shi for technical contributions.
\end{acknowledgments}

%\nocite{*}  % show all references including the uncited ones

\bibliography{main}% Produces the bibliography via BibTeX.

\end{document}